\documentclass[aps,pra,twocolumn,showpacs,amsmath,amssymb]{revtex4-1}
\usepackage{graphicx}

\usepackage{natbib}

\usepackage{xspace}
\newcommand{\rcm}{\mbox{cm$^{-1}$}\xspace}
\newcommand{\Xstate}{X$^1\Sigma^+$\xspace}
\newcommand{\Bstate}{B$^1\Pi$\xspace}

\begin{document}


\title{The permanent dipole moment of LiCs in the ground state}

\author{J. Deiglmayr}
\altaffiliation{Permanent address: Albert-Ludwigs-Universit\"at Freiburg, Physikalisches
Institut, Hermann-Herder-Str. 3, 79104 Freiburg, Germany}
\author{A. Grochola}
\altaffiliation{Permanent address: Institute of Experimental Physics,
Warsaw University, Poland}
\author{M. Repp}
\author{O. Dulieu}
\altaffiliation{Permanent address: Laboratoire Aim\'e Cotton, CNRS, Universit\'e Paris-Sud XI, Orsay, France}
\author{R. Wester}
\altaffiliation{Permanent address: Albert-Ludwigs-Universit\"at Freiburg, Physikalisches
Institut, Hermann-Herder-Str. 3, 79104 Freiburg, Germany}
\author{M. Weidem\"uller}
\email{weidemueller@physi.uni-heidelberg.de}
\affiliation{Ruprecht-Karls-Universit\"at Heidelberg, Physikalisches
Institut, Philosophenweg 12, 69120 Heidelberg, Germany}

\begin{abstract}
Recently we demonstrated the formation of ultracold polar LiCs molecules in deeply bound levels of the \Xstate ground state, including the rovibrational ground state~[J. Deiglmayr \textit{et al.}, Phys. Rev. Lett. 101, 133004 (2008)]. Here we report on the first experimental determination of the permanent electric dipole moment of deeply bound LiCs molecules. For \Xstate,$v''$=2 and $v''$=3 we measure values of $\mu$=5.5(2)\,Debye and 5.3(2)\,Debye respectively.
\end{abstract}

\date{\today}



\maketitle


  Ultracold gases of dipolar molecules have been proposed as candidates for the exploration of quantum phases in dipolar gases\cite{Pupillo2008b}, the development of quantum computation techniques\cite{DeMille2002}, or precision measurements of fundamental constants~\cite{zelevinsky2008}. LiCs is predicted to have the largest permanent electric dipole moment (EDM) of all alkali dimers~\cite{Aymar2005} and is thus very advantageous for such schemes. Recently we have reported the formation of LiCs molecules in deeply bound levels of the \Xstate ground state~\cite{Deiglmayr2009a}, including the rovibrational ground state~\cite{Deiglmayr2008b}, after a single step of PA~\cite{Deiglmayr2008b}. Here we report on the first experimental determination of the permanent EDM of deeply bound LiCs molecules.

  \begin{figure}[htb]
   \begin{center}
   \includegraphics[width=\columnwidth,clip]{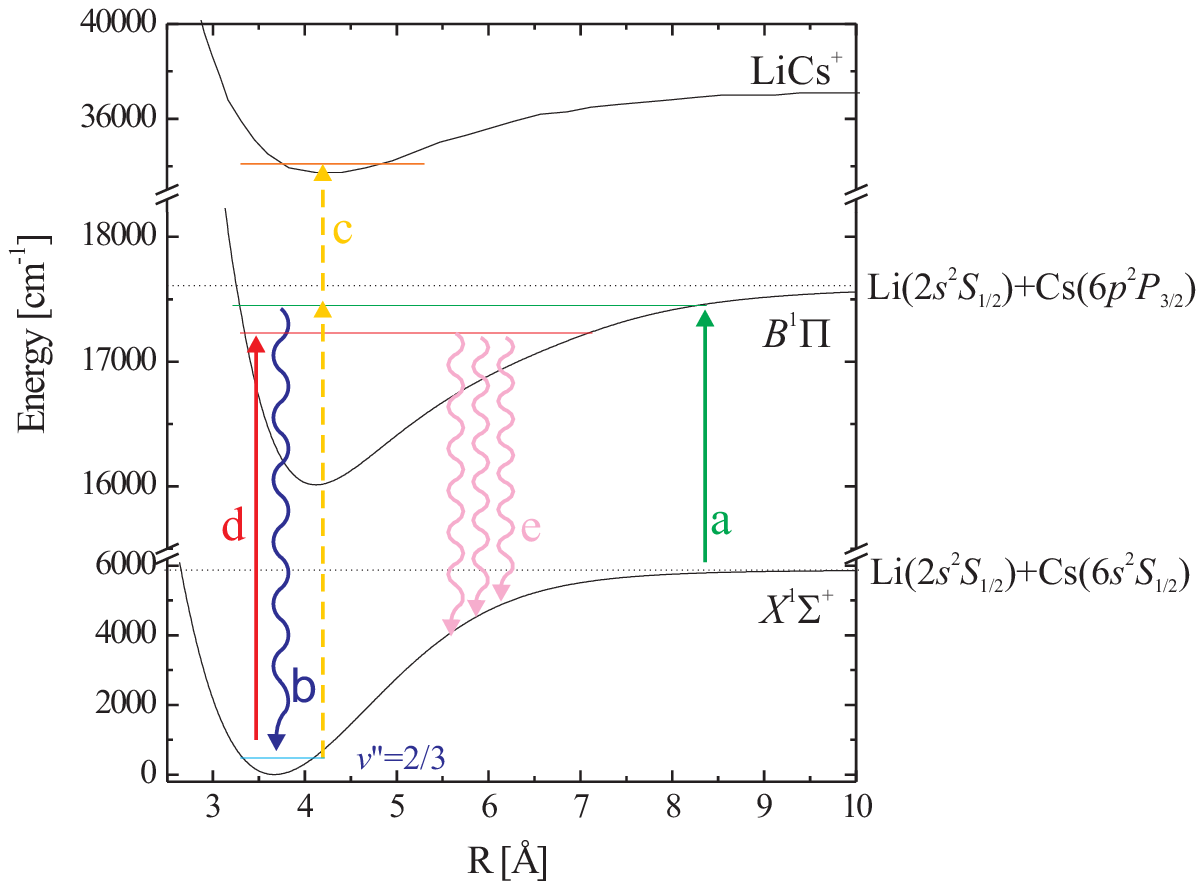}
   \end{center}
   \caption{Sketch of the excitation and detection scheme. a) photo-association, b) spontaneous decay into deeply bound ground state molecules, c) two photon ionization with resonant intermediate state; for depletion spectroscopy: d) excitation of ground state molecules, e) redistribution of ground state population}
   \label{fig:levelScheme}
  \end{figure}

  The experimental sequence for formation, detection, and high-resolution spectroscopy of ground state molecules is shown in fig.~\ref{fig:levelScheme}. Ultracold LiCs molecules are formed by photoassociation (PA)~\cite{thorsheim1987,jones2006} of laser cooled $^{133}\,$Cs and $^{7}$Li atoms. The atoms are cooled and trapped in two overlapped magneto-optical traps (MOTs) which are loaded via a single Zeeman slower from a double species oven. For the Cs atoms the MOT is operated in the configuration of a dark spontaneous force optical trap (dark SPOT)~\cite{ketterle1993} to reduce inelastic inter-species collisions. In this configuration $4\times10^7$ $^{133}\,$Cs and 10$^8$ $^{7}$Li atoms are trapped simultaneously at densities of $3\times10^9\,$cm$^{-3}$ and $10^{10}\,$cm$^{-3}$, and temperatures of 250(50)$\mu$K and 600(150)$\mu$K respectively. Details of the experimental setup for the formation and detection of ultracold LiCs molecules have been given elsewhere~\cite{Deiglmayr2008b,Deiglmayr2009a,Kraft2007}. For PA of LiCs molecules the two overlapped atom clouds are continuously illuminated by light from a tunable cw TiSa laser (typically 500\,mW, collimated to 1.0\,mm). The frequency of the laser is stabilized to a transfer cavity which is in turn stabilized via a frequency-offset locked diode laser ~\cite{Schunemann1998a} to an atomic cesium resonance. This locking scheme results in residual shifts of less than 1\,MHz for the central PA frequency. The excited LiCs molecules decay spontaneously within few tens of nanoseconds into different electronic ground state levels. Molecules in given vibrational ground state levels are then ionized by resonant-enhanced multi-photon ionization (REMPI) and the resulting ions are detected in a time-of-flight mass spectrometer~\cite{Kraft2007}.

  Rotationally resolved spectroscopy of ground state molecules is achieved by depletion spectroscopy~\cite{Wang2007}: a narrow band laser pumps population out of a selected rovibrational ground state level leading to a reduction in the detected ion signal from this level. Further details have been given previously~\cite{Deiglmayr2009a}. The light for depumping is provided by a cw dye laser and transferred to the experimental setup by a single-mode optical fiber. Here the collimated beam with a waist of 0.7\,mm is overlapped and aligned colinear with the PA light using a selectively coated mirror. The intensity of the depumping light after the fiber is monitored using a beam sampler and a photodiode (PD). The PD signal is fed back to an electro-optical modulator in front of the fiber to compensate intensity fluctuations originating from the laser and the fiber coupling. The frequency of the cw dye laser is stabilized via a transfer cavity to an atomic cesium resonance. Using an AOM in the locking branch of the dye laser, relative frequency shifts of few hundred MHz can be controlled with an accuracy of few MHz. In the depletion spectroscopy we reach a resolution of typically 5\,MHz for shifts in transition frequencies, mostly limited by the hyperfine-induced broadening of the excited level used for depletion, and an absolute accuracy of 500\,MHz, given by the resolution of the employed wavemeter (Burleigh WA-1000).

  Electric fields between 0 and 500\,V/cm can be applied to the atomic and molecular samples by two round electrodes inside the vacuum with an outer diameter of 80\,mm and a distance of 40\,mm, centered around the trapped atom clouds. The electrodes have a central hole ({\o}30\,mm) to allow for optical access and extraction of produced ions (these electrodes are also part of the mass-spectrometer used for the detection of the ground state molecules~\cite{Kraft2007}). The electric field created at the trap position is derived from the measured dimensions and voltages using the simulation package SIMION (v8.0) with an uncertainty of 3\% due to estimated inaccuracies in the electrode geometry and voltages. For the detection of molecules the fields are switched 0.6\,ms before ionization to a fixed value of 38\,V/cm.

  In order to determine the permanent EDM of a given vibrational level of the \Xstate state we measure the electric-field induced shift in the transition frequency between the $J''$=0 component of this level and a rovibrational level in the \Bstate state. A shift in the transition frequency originates in the combined Stark shifts of the \Xstate level and the excited \Bstate level. A rotationless level $v'$ ($J''$=0) of the \Xstate state experiences an electric-field induced shift
  \begin{equation}
  \vartriangle W=-{\mu_v^2 E^2}/{(6 h B_v)},
  \label{eq:starkeffect}
  \end{equation}
  where $\mu_v$ is the permanent EDM, $E$ the external field, and $B_v$ the rotational constant~\citep{Town1955}. Rotational constants are calculated from an experimental \Xstate potential energy curve~\cite{Staanum2007} using Level 8.0~\cite{Leroy2007}. Due to selection rules the upper level in the depletion transition has to be rotationally excited ($J'$=1). As the \Bstate state is also predicted to have a smaller EDM than the ground state~\footnote{A pseudopotential configuration interaction calculation as described in ref.~\cite{Aymar2005} yields for the \Bstate state a permanent EDM of 0.47\,Debye at the equilibrium distance.} one can expect a significantly smaller Stark effect for the upper level than for the ground state level. In order to determine this experimentally we perform PA spectroscopy of the \Bstate,$v'$=18,$J'$=1 level~\cite{Grochola2009}, the upper level of the employed depletion transitions, at different electric fields. As the initial state of the PA process is a free pair of atoms without permanent EDM, every change in the line shape and position is exclusively due to the Stark effect of the excited level. The width of the PA resonance increased from 50\,MHz FWHM at zero field\footnote{Due to thermal broadening the PA resonances are significantly broader than the depletion resonances~\cite{Deiglmayr2009a}, which were found to be around 30\,MHz at zero field.} to 90\,MHz FWHM at the largest applied field of 488.0 V/cm. However the line center did not experience significant shifts. Thus any change in the transition frequency of a depletion resonance can be exclusively attributed to the Stark shift of the ground state level.

\begin{figure}[b]
   \includegraphics[width=0.95\columnwidth,clip]{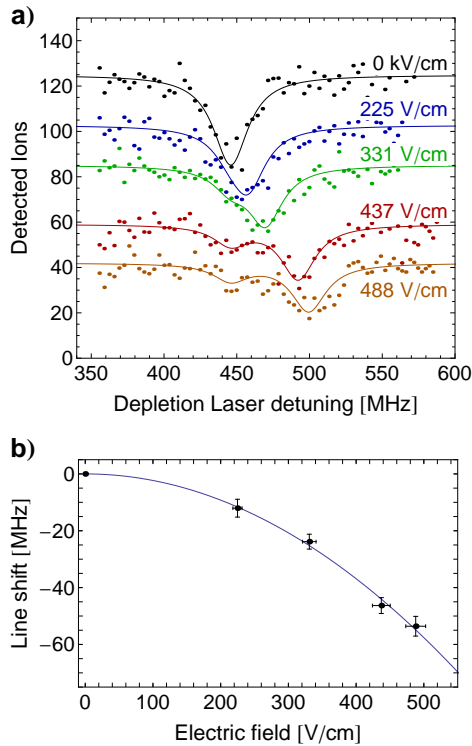}
   \caption{Stark shift of the \Xstate,$v''$=2,$J''$=0$\rightarrow$\Bstate,$v'$=18,$J'$=1 transition: a) depletion scans at different electric fields (dots, lower traces offset by 20 counts each for clarity) and Lorentzian fits to the data (solid lines). The electric field is switched 0.6\,ms before detection to a fixed value of 38\,V/cm for the extraction of ions. Molecules which are depleted during this period experience a negligible Stark shift and lead to an unshifted feature in each scan, which is included in the fit. Each trace is an average over 4 scans; b) fitted resonance positions (symbols) together with a fit to a quadratic Stark effect (solid line).}
   \label{fig:starkv2}
\end{figure}

  For the formation of deeply bound molecules we use the photoassociation resonance \Bstate,$v'$=18,$J'$=1. After spontaneous decay the rotational states $J''$=0 of \Xstate,$v''$=2 and higher are significantly populated. Then depletion spectroscopy of $v''$=2,$J''$=0 and $v''$=3,$J''$=0 is performed in electric fields varying from 0 to 488\,V/cm. We did not obtain depletion spectra with sufficient signal to noise ratio for Stark spectroscopy from the $J''$=0 component of the $v''$=0 and $v''$=1 levels~\footnote{Molecules in the lowest vibrational levels $v''$=0 and $v''$=1 are also formed by photoassociation, but mostly in the rotationally excited $J''$=2 components~\cite{Deiglmayr2008b,Deiglmayr2009b} for which the Stark effect at feasible electric fields could not be experimentally observed, yet.}. The change in the frequency of the transition \Xstate,$v''$=2,$J''$=0$\rightarrow$\Bstate,$v'$=18,$J'$=1 is determined via depletion spectroscopy, as shown in fig.~\ref{fig:starkv2}\,a), where \Xstate,$v''$=2 molecules are selectively detected via the intermediate state \Bstate,$v'$=19 at a REMPI wavelength of 16\,860.0\rcm. The position of the depletion resonance is obtained from a fit of two Lorentzian line profiles to the spectrum, one for the shifted resonance and one for an unshifted feature, which is introduced by switching the fields to a small value during detection of molecules.

  Fitting eq.~(\ref{eq:starkeffect}) to the data and using $B_{v''=2}$=0.1848\,\rcm we find a permanent EDM of $\mu$=5.5(2)\,Debye for $v''$=2 (see fig.~\ref{fig:starkv2}\,b)), where the reported error stems from the uncertainty of the fit and the estimated uncertainty on the electric field determination. The same measurement was performed for molecules in \Xstate,$v''$=3,$J''$=0. These molecules are detected at a REMPI frequency of 16\,886.3\rcm (the intermediate level is a vibrational level in a $\Omega$=0 state, probably associated to $A^1\Sigma^+$). Using $B_{v''=3}$=0.1836\,\rcm the value for the EDM is determined as $\mu$=5.3(2)\,Debye.

  While this is the first experimental determination of the permanent EDM of LiCs in the ground state, few results of \textit{ab-initio} calculations have been published. Igel-Mann \textit{et al.} perform pseudopotential configuration interaction calculations and report a vibrational average for \Xstate,$v''$=0 of 5.48\,Debye~\cite{Igel-Mann1986}. Aymar and Dulieu use a similar approach with refined basis sets and obtain a value of~5.478\,Debye~\cite{Aymar2005}. S{\o}rensen \textit{et al.} employ a coupled cluster model with full iterative triple excitations correlating 10 electrons and report a value of 5.440\,Debye~\cite{Sorensen2009}. Only Ref.~\cite{Aymar2005} also shows vibrational averages for higher vibrational levels. Due to the coincidence of the equilibrium distance of the \Xstate state potential with a linear part of the dipole moment curve~\cite{Sorensen2009} the dependence of the EDM on the vibrational level is only weak. For the here considered levels $v''$=2 and $v''$=3 the values are found to be at most 1\% larger than the value for $v''$=0. Therefore we would not expect a significantly different measurement for the dipole moment of the $v''$=0 ground state.

  The large permanent EDM of deeply bound LiCs makes this molecule a promising candidate for the realization of an ultracold gas with strong dipolar interactions. However a permanent EDM also couples the internal molecular state to the thermal environment via black-body radiation (BBR) and thus leads to a finite lifetime even for molecules in the absolute ground state~\cite{Vanhaecke2007,Buhmann2008}. As a second consequence of the existence of a permanent EDM, spontaneous decay of electronic ground state molecules into more deeply bound rovibrational levels is possible. In the case of LiCs the timescale for such processes is expected to be in the range of few tens of seconds and should thus be observable during the lifetime of the molecules in an optical dipole trap. We will discuss the importance of these processes in a future publication.

  \begin{acknowledgments}
  This work is supported by the DFG under WE2661/6-1 in the framework of the Collaborative Research Project QuDipMol within the ESF EUROCORES EuroQUAM program. JD acknowledges partial support of the French-German University. AG is a postdoctoral fellow of the Alexander von Humboldt-Foundation.
\end{acknowledgments}

\end{document}